\documentclass[aps,pra,superscriptaddress,tightenlines,showpacs]{revtex4}
\usepackage{graphicx} 
\usepackage{amsmath}
 
\newcommand{\beq}{\begin{equation}}
\newcommand{\enq}{\end{equation}}

\begin{document}


\title{Spontaneous squeezing of a vortex in an optical lattice} 
\author{J.-P. Martikainen}
\author{H. T. C. Stoof}
\affiliation{Institute for Theoretical Physics, Utrecht University, 
Leuvenlaan 4, 3584 CE Utrecht, The Netherlands}
\date{\today}

\begin{abstract}
We study the equilibrium states of a vortex in a Bose-Einstein condensate in 
a one-dimensional optical lattice. We find that quantum effects
can be important and that it is even possible for the vortex
to be strongly squeezed, which reflects itself in a 
different quantum mechanical uncertainty of the vortex 
position in two orthogonal directions. The latter is observable by measuring
the atomic density after an expansion of the Bose-Einstein condensate in the lattice.
\end{abstract}
\pacs{03.75.-b, 32.80.Pj, 03.65.-w}  

\maketitle

\section{Introduction}
\label{sec:intro}
Vortices play a crucial role in explaining rotational and dissipative properties of superfluids. 
In the superfluid systems studied until now, such as liquid Helium~\cite{Donnelly}, 
superconductors~\cite{Blatter1994a}, and 
Bose-Einstein condensates~\cite{Matthews1999b,Madison2000a,Abo-Shaeer2001a,Hodby2002a,
Butts1999a,Williams1999a,Dodd1997a,Zambelli1998a,Fetter2001a}, 
a vortex line behaves as a classical object. However, 
placing a Bose-Einstein condensate in an optical 
lattice~\cite{Cataliotti2001a,Burger2001a,Orzel2001a,Greiner2002a,Greiner2002b} 
leads to an unprecedented control over the system 
parameters, which has already enabled experimental studies on quantum 
phase-transitions~\cite{Greiner2002a,Jaksch1998a,vanOosten2001a}, 
superfluidity~\cite{Cataliotti2001a,Burger2001a}, and number squeezing~\cite{Orzel2001a}. 
Furthermore, 
a Bose-Einstein condensate in a one-dimensional optical lattice has a similar layered structure 
as the cuprate superconductors and is a promising way to achieve the quantum Hall regime in a 
Bose-Einstein condensed gas~\cite{Cornell2002private}. 
This suggests that a careful examination of the quantum 
properties of a vortex line in a one-dimensional optical lattice is warranted.  

	A characteristic feature of all superfluids is their ability to support persistent 
currents. Quantized vortices play a crucial role in understanding the decay of these currents, 
and also the superfluid's response to rotation or to an external magnetic field in the case of 
superconductor. In pioneering work Fetter developed the quantum theory of vortices in liquid 
Helium~\cite{Fetter1967a} and so-called type-${\rm II}$ superconductors~\cite{Fetter1967b}. 
In these systems quantum fluctuations turned 
out to play such a small role that they are experimentally inaccessible. Here we show, 
however, that truly quantum mechanical behavior of the macroscopic vortex occurs in a 
Bose-Einstein condensate in a one-dimensional optical lattice. This is due to the reduced 
dimensionality as well as the reduced number of atoms in every well of the optical lattice. 
Remarkably, it turns out that the vortex can spontaneously become strongly squeezed, which is reflected 
in the quantum-mechanical probability distribution of the vortex position.
Unlike coherent states, which are described by Gross-Pitaevskii theory,
squeezed states are highly nonclassical.

\section{Bose-Hubbard Hamiltonian}
We consider a Bose-Einstein condensate in a one-dimensional optical lattice with lattice 
spacing $\lambda/2$, where $\lambda$ is the wavelength of the laser beams 
creating the standing wave of the lattice. The optical lattice splits the 
Bose-Einstein condensate into a stack of $N_s$ weakly-coupled pancake condensates, each 
containing $N$ atoms. For concreteness we use in the following always $^{87}{\rm Rb}$ atoms. The 
depth of the lattice $V_L$  
can be easily changed and controls the tunneling of an atom from 
one pancake condensate to the next, i.e., the strength of the interlayer Josephson coupling. 
The condensate also experiences a harmonic trapping potential in the radial direction 
with an oscillator frequency $\omega_r$. The longitudinal trapping potential along the direction 
of the laser beams is assumed to be so weak that it can be neglected.  While the lattice 
is taken to be deep enough to allow us to use a tight-binding approximation and to 
include only the weak nearest-neighbor Josephson coupling, it is also taken to be shallow 
enough to support a superfluid state as opposed to the Mott-insulator state~\cite{vanOosten2001a}.

Furthermore, we consider a vortex line that pierces at the position 
$\left(x_n,y_n\right)$ through each 
layer of the stack labelled by the index $n$. 
In each layer the density of the pancake condensate 
varies as a function of the radial distance. As a result 
the vortex experiences an effective 
potential that depends on its distance from the origin~\cite{Fetter2001a}. 
Without rotation of the gas this 
potential is approximately an inverted parabola. Hence, 
the vortex is energetically unstable 
and will tend to spiral out of the system. However, 
the vortex is stabilized if we rotate the 
Bose-Einstein condensate with a rotation frequency $\Omega$ 
that is larger than a critical frequency $\Omega_c$. In the 
rest of this paper we consider the regime where the vortex 
is energetically stable and $\Omega>\Omega_c$. 

In the tight-binding approximation the attraction between 
the nearest-neighbor parts of the vortex 
turns out to be harmonic with respect to their separation. 
We denote the typical strength of this attraction by $J_V$. 
Physically the attractive interaction is due to the energy 
cost for phase differences between 
the two layers. The theory of the resulting coupled harmonic 
oscillators can be quantized by 
introducing the bosonic annihilation operators for 
the eigenmodes of the vortex 
line~\cite{Bretin2003a,Mizushima2003a,Martikainen2003b}. 
These modes are the Kelvin modes (or kelvons) and
correspond physically to a wiggling of the vortex 
line with a wavelength $2\pi/k$. 
The operators $\hat{v}_k$ in momentum space are related 
to the coordinate space kelvon operators $\hat{v}_n$ 
and to the vortex positions through  
\beq
\hat{x}_n=\frac{R}{2\sqrt{N}}(\hat{v}_n^\dagger+\hat{v}_n)=\frac{R}{2\sqrt{NN_s}}
\sum_k e^{ikn\lambda/2}\left(\hat{v}_{-k}^\dagger+\hat{v}_k\right)
\enq
and  
\beq
\hat{y}_n=\frac{iR}{2\sqrt{N}}(\hat{v}_n^\dagger-\hat{v}_n),
\enq
where $R$ is the typical radial 
size of each pancake condensate. In this way we obtain the dispersion relation 
$\hbar\omega_K(k)=
\hbar\omega_K(0)+J_V\left[1-\cos\left(k\lambda/2\right)\right]$ of 
the kelvons~\cite{Martikainen2003b}. 
The energetic stability of the vortex is 
reflected in a positive value of the dispersion at 
zero momentum. Physically the frequency at zero momentum 
is the precession frequency of 
a slightly displaced straight vortex line around 
the center of the Bose-Einstein condensate. 

It is crucial for our purposes to realize that, due to the Euler dynamics of the vortex, 
the coordinates $\hat{x}_n$ and $\hat{y}_n$ 
are canonically conjugate variables and therefore obey the Heisenberg 
uncertainty relation $\left[\hat{x}_n,\hat{y}_n\right]=iR^2/2N$. 
This means that the position of the vortex is always smeared out by 
quantum fluctuations. For a vortex in a lattice these fluctuations can become large due to 
the reduced particle numbers in every layer, as opposed to the total number of particles, 
and the radial spreading of the pancake condensate wave function as the lattice depth is increased.  

The kelvon dispersion relation was determined by taking into account only the lowest 
order expansion of the effective potential experienced by the vortex. However, expanding up to 
fourth order in the vortex displacements results in an interaction 
$\left(V_0/2\right)\sum_n\hat{v}_n^\dagger\hat{v}_n^\dagger\hat{v}_n\hat{v}_n$
for the kelvons. 
In the regime where the vortex is stable $V_0<0$ and this interaction is attractive. 
The physics of 
the vortex line is thus described by a one-dimensional Bose-Hubbard model with a negative 
interaction strength. The corresponding Hamiltonian is given by
\beq
\hat{H}=\sum_n\left[\hbar\omega_K(0)+J_V\right]\hat{v}_n^\dagger\hat{v}_n
-\frac{J_V}{2}\sum_{\langle n,m\rangle}\hat{v}_m^\dagger\hat{v}_n
+\frac{V_0}{2}\sum_n\hat{v}_n^\dagger\hat{v}_n^\dagger\hat{v}_n\hat{v}_n,
\label{eq:H}
\enq
where $\langle n,m\rangle$ denotes the nearest-neighbor layers.

We calculate the parameters of this Bose-Hubbard model using 
a variational ansatz we used in our earlier work~\cite{Martikainen2003b},
where the density of the condensate wave function in each layer
was proportional to $\exp\left[-\left(x^2+y^2\right)/R^2\right]$.
In this way we find that the strength of the nearest-neighbor coupling is
given by
\begin{equation}
J_V=\frac{\hbar\omega_r}{4\pi^2}\Gamma\left[0,\frac{l_r^4}{R^4}\right]
\left(\frac{\omega_L\lambda}{\omega_r l_r}\right)^2
\left(\frac{\pi^2}{4}-1\right)\exp\left(-\frac{\lambda^2 m\omega_L}{4\hbar}\right)
\nonumber
\label{eq:J}
\end{equation}
and the interaction strength  is given by
\beq
V_0=\frac{2\hbar\omega_r\left(l_r/R\right)^2\Gamma\left[0,l_r^4/R^4\right]-3\hbar\omega_r
\left(l_r/R\right)^2-4\hbar\Omega}{4N}\nonumber,
\label{eq:V0}
\enq
where $\omega_L=\sqrt{8\pi^2V_L/m\lambda^2}$ is the 
oscillator frequency of the optical lattice, $l_r=\sqrt{\hbar/m\omega_r}$,
and $\Gamma\left[a,z\right]$ is the incomplete gamma function~\cite{comment2004a}.
Furthermore, a straight and slightly displaced vortex precesses around 
the condensate center with 
the frequency 
\beq
\omega_K(k=0)=\left(\omega_r l_r^2/2R^2\right)
\left(1-\Gamma\left[0,l_r^4/R^4\right]\right)+\Omega.
\enq 
The precession frequency of the vortex changes from negative to positive
at the critical rotation frequency
\beq
\Omega_c=\left(\omega_r l_r^2/2R^2\right)
\left(\Gamma\left[0,l_r^4/R^4\right]-1\right).
\enq
When $\Omega>\Omega_c$ the vortex is locally energetically stable.
The condensate size
$R$ is given by the solution of the transcendental equation 
$\left(l_r/R\right)^4
\left[1-\gamma+2Na\sqrt{m\omega_L/2\pi\hbar}-4\ln\left(l_r/R\right)\right]=1$, 
where $\gamma$ is the Euler-Mascheroni constant and $a$ is the three-dimensional
scattering length.
Without the kelvon interaction the ground state of the vortex 
corresponds to a vortex line along 
the symmetry axis of the Bose-Einstein condensate and with a 
quantum mechanical uncertainty such that 
$\langle \hat{x}_n^2\rangle=\langle \hat{y}_n^2\rangle$. 
An attractive interaction will lead to squeezing with  
$\langle \hat{x}_n^2\rangle\neq\langle \hat{y}_n^2\rangle$ 
as we show next.

\section{Vortex squeezing}
First we have to identify the correct order parameter for the vortex line. 
Since $\Omega>\Omega_c$ there occurs 
no Bose-Einstein condensation of the kelvons and the appropriate order parameter for 
the vortex line is not $\langle \hat{v}_n\rangle$, 
which would signal the above mentioned energetic instability and 
the tendency of the vortex to move away from the symmetry axis. 
However, the correct order 
parameter can be identified with $\Delta=V_0\langle \hat{v}_n\hat{v}_n\rangle$. 
To arrive at the associated mean-field theory, 
we quadratically expand the Hamiltonian around this order parameter and 
obtain apart from a constant~\cite{Stoof1994a}
\beq
\hat{H}=\sum_n\left[\hbar\omega_K(0)+J_V\right]\hat{v}_n^\dagger\hat{v}_n
-\frac{J_V}{2}\sum_{\langle n,m\rangle}\hat{v}_m^\dagger\hat{v}_n
+\sum_n\left[\frac{\Delta}{2}\hat{v}_n^\dagger\hat{v}_n^\dagger
+\frac{\Delta^*}{2}\hat{v}_n\hat{v}_n\right].
\label{eq:meanH}
\enq
We diagonalize this result by means of a Bogoliubov transformation. 
It is most convenient to work in the momentum space where
we define new creation and annihilation operators with the help
of of the Bogoliubov amplitudes $u_k$ and $v_k$ as
$\hat{b}_k=u_k\hat{v}_k^\dagger-v_k\hat{v}_{-k}$
and $\hat{b}_k^\dagger=u_k^*\hat{v}_k-v_k^*\hat{v}_{-k}^\dagger$.
Furthermore, to ensure bosonic commutation relations we must have $|u_k|^2-|v_k|^2=1$.
The Hamiltonian in Eq.~(\ref{eq:meanH}) is then diagonalized with the choise
\beq
|v_k|^2=\frac{|\Delta|^2}{\left(E(k)+\omega_K(k)\right)^2-|\Delta|^2}
\enq
and 
\beq
|u_k|^2=\frac{\left(E(k)+\omega_K(k)\right)^2}
{\left(E(k)+\omega_K(k)\right)^2-|\Delta|^2}, 
\enq
where $E(k)=\sqrt{\omega_K(k)^2-|\Delta|^2}$ is the dispersion 
of the Bogoliubov quasiparticles.

\begin{figure}
\includegraphics[width=\columnwidth]{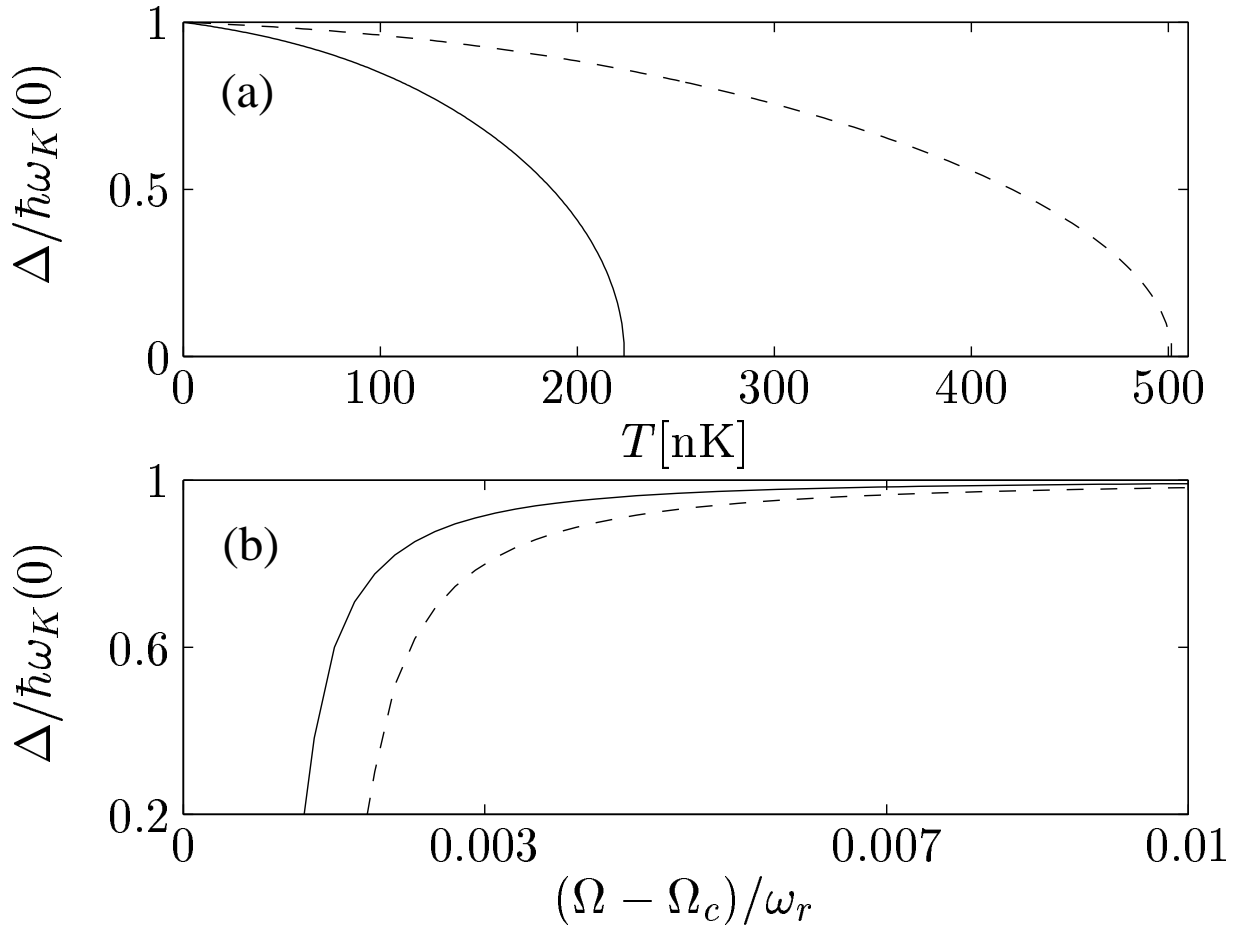}
\vspace{-0.5cm}
\caption[Fig1]{Solution of the gap equation as a function of temperature and rotation
frequency.\,\,\,(a) Shows the order parameter as a function of temperature for two
different rotation frequencies. The solid line represents $\Omega-\Omega_c=0.01\cdot\omega_r$
and the dashed line represents $\Omega-\Omega_c=0.02\cdot\omega_r$. The result 
for the order parameter is scaled to the precession frequency of the vortex.
(b) Shows the order parameter as a function of rotation frequency at two different
temperatures. The solid line represents a temperature $T=10\, {\rm nK}$ and the dashed
line represents the temperature $T=20\, {\rm nK}$. The curves are calculated
for the parameters $\lambda=795\, {\rm nm}$, $\omega_r=2\pi\cdot 100\, {\rm Hz}$,
$N=1000$, $N_s=51$, and $V_L=15\,E_r$, where $E_r=2\pi\hbar^2/m\lambda^2$ is the
recoil energy of the atom after absorption of a photon from the laser beam.
These parameters are representative of current
experimental capabilities. We use this same set of parameters also in the other 
figures.}
\label{fig:gapeq}
\end{figure}

Requiring self-consistency of 
the approach leads to a gap equation
\beq
\frac{1}{V_0}=-\frac{1}{N_s}\sum_k\frac{1+2N_k}{2E(k)},
\label{eq:gap}
\enq
where $N_k=1/(e^{\beta E(k)}-1)$ is the Bose distribution.
We solve the gap equation for the order parameter, 
analogous to the BCS theory of superconductors~\cite{Schrieffer1964a}. In Fig.~\ref{fig:gapeq} 
we present the behavior of 
the order parameter as a function of temperature and rotation frequency for typical 
parameter values. We see from Fig.~\ref{fig:gapeq}
that there is a critical temperature below which 
the order parameter becomes nonzero and then increases monotonically with decreasing 
temperature. Furthermore, the absolute value of the order parameter at zero 
temperature is usually close to the precession frequency. Only very close 
to the critical rotation frequency $\Omega_c$ the deviation becomes large.

\begin{figure}
\includegraphics[width=0.97\columnwidth]{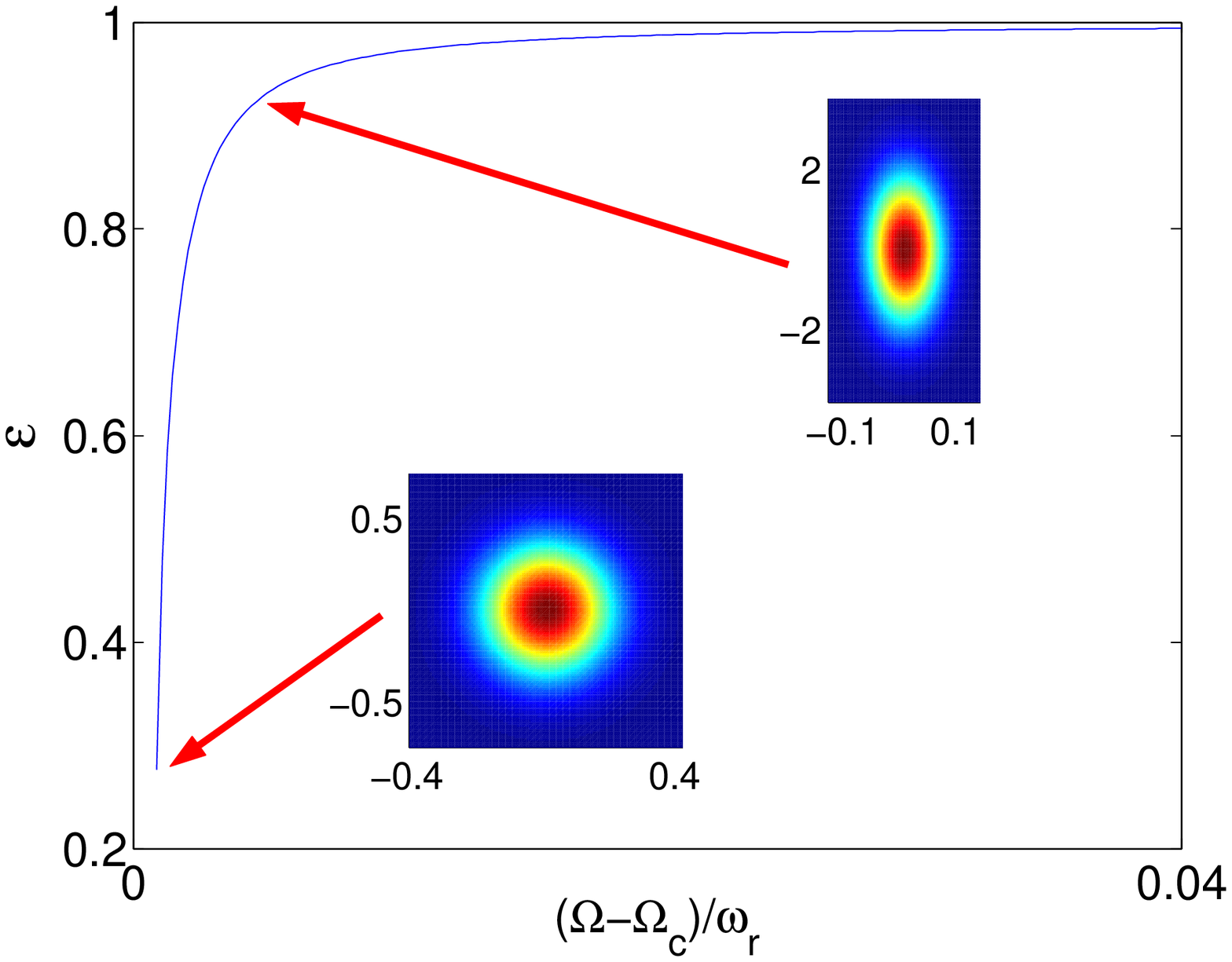}
\vspace{-0.3cm}
\caption[Fig2]{(Color online) Squeezing of the quantum mechanical position uncertainty ellipse.
The figure shows the relative squeezing 
$\epsilon=\langle \hat{y}^2-\hat{x}^2\rangle/\langle \hat{x}^2+\hat{y}^2\rangle$
of the uncertainty ellipse of the vortex position as a function of rotation frequency
at a low temperature of $5\, {\rm nK}$. Subplots show samples of the Gaussian 
probability distributions for the vortex position for two different rotation 
frequencies. In the subplots we use the trap length 
$l_r=\sqrt{\hbar/m\omega_r}$ as a 
unit of length. In order to ease visual inspection we use different aspect ratios for the
axes of the subplots. For the parameters used the coherence length
at the center of the condensate is about $0.34\, l_r$.
}
\label{fig:squeeze}
\end{figure}

The complex order parameter $\Delta=|\Delta| e^{i\phi}$ 
has an interesting physical interpretation in terms of the quantum 
mechanical uncertainty of the vortex position. It turns out that in a coordinate system rotated 
by an angle $\theta$ we have 
\beq
\langle \hat{y}_n^2\rangle-\langle \hat{x}_n^2\rangle=
\frac{|\Delta|R^2\cos\left(\phi-2\theta\right)}{|V_0|NN_s},
\enq 
where $\phi$ is the phase of the order parameter. This implies that a nonzero 
value of the order parameter is reflected in the squeezing of the vortex position distribution. 
The main axis of the uncertainty ellipse is at an angle $\theta=\phi/2$ with the x-axis. 
In equilibrium 
the uncertainty ellipse of the vortex position distribution is independent of 
the layer index $n$. 
Therefore, the measurement of the vortex positions in different layers samples the same 
distribution and provides a signature for the expected transition into a squeezed state. 
In Fig.~\ref{fig:squeeze} 
we plot the relative squeezing as a function of rotation frequency at fixed temperature. 
As can be seen from this figure the size of the uncertainty ellipse can easily be several times 
the radial trap length and also becomes very strongly deformed as the 
rotation frequency is increased. 
Furthermore, the long axis of the uncertainty 
ellipse can be much larger than the vortex core size. 
The results in Fig.~\ref{fig:squeeze} were obtained with only one set of typical parameters. 
By a careful choice of, for example, lattice depth, number of particles, number of layers, 
or temperature a considerable degree of tunability is possible. For example, a shallower 
lattice implies a larger relative squeezing.

\section{Phase fluctuations}
	For our theory to be valid, constraints have to be set for the temperature. 
Since we assume a pure condensate, we are ignoring the influence of the noncondensate 
atoms on the vortex line. Therefore, the temperature should be much lower than the critical 
temperature for Bose-Einstein condensation in the layers. This condition is relatively 
easy to satisfy. However, a more stringent condition is set by the phase fluctuations of 
the order parameter. In an infinite one-dimensional system phase fluctuations destroy 
the long-range order. In the finite system we are considering here, phase fluctuations 
can only be excited if the temperature is high 
enough~\cite{Petrov2000a,Andersen2002a,Khawaja2002a}.  
In order to calculate 
the energy cost for a phase gradient of the order parameter we must determine the 
associated stiffness or superfluid density $\rho_S(T)$. Since the lattice breaks the Galilean 
invariance this calculation is not entirely standard. 

The superfluid density is defined by the systems response to an imposed phase 
gradient~\cite{Rey2003a}. 
Therefore, we want to study the 
long-wavelength effects of the transformation 
$\hat{v}_n\rightarrow \hat{v}_ne^{i\phi_n/2}$. 
The phase gradients enter only in the 
nearest-neighbor coupling and result in an interaction energy 
\begin{equation}
\hat{H}_I=-\frac{J_V}{2}\sum_{\langle n,m\rangle}\hat{v}_m^\dagger\hat{v}_n\left[
i\left(\phi_n-\phi_m\right)/2-\left(\phi_n-\phi_m\right)^2/8\right]\nonumber,
\label{eq:HI}
\end{equation}
which can be considered as a disturbance in the long-wavelength limit. 
With this interaction energy we calculate in second-order
perturbation theory the contribution to the free energy of the 
phase gradients. 
Up to second order the effective imaginary time action for the phase fluctuations is
\beq
S_\phi=\int_0^{\hbar\beta}d\tau\langle H_I\left(\tau\right)\rangle
-\frac{1}{2\hbar}\int_0^{\hbar\beta}\int_0^{\hbar\beta}d\tau d\tau'
\langle H_I\left(\tau\right)H_I\left(\tau'\right)\rangle.\nonumber
\label{eq:Sphi}
\enq
Furthermore, it is sufficient to keep only the terms up to second order in the phase 
differences between layers. In this way we find the action for the phase gradients as
\beq
S_\phi=\int_0^{\hbar\beta} d\tau \frac{J_V}{8N_s}
\sum_{k'}\left\{\cos\left(k'\lambda/2\right) \left[|u_{k'}|^2N_{k'}+|v_{k'}|^2
\left(N_{k'}+1\right)\right]
-J_V\beta \sin^2\left(k'\lambda/2\right)N_{k'}\left(N_{k'}+1\right)
\right\}\times\sum_k k^2 \phi_k\phi_{-k}.
\label{eq:phaseaction}
\enq
This result should be compared with the standard (Galilean invariant)
Landau~\cite{Schrieffer1964a} result
\beq
S_L=\int_0^{\hbar\beta}d\tau
\int dz \frac{\hbar^2}{2m_\Delta}\rho_s(T)|\nabla\phi|^2,
\enq
where the superfluid density at temperature $T$ is
\beq
\rho_s(T)=\frac{1}{N_s\lambda}\sum_k 
\left[|u_k|^2N_k+|v_k|^2\left(N_k+1\right)-\frac{\hbar^2\beta}{m_\Delta}
k^2N_k\left(N_k+1\right)\right].
\enq
When we take the lattice spacing to zero while keeping
$J_V\lambda^2$ constant, we can see that Eq.~(\ref{eq:phaseaction})
recovers the standard result
when the mass of the pairing field is defined as $m_\Delta=2m_K$, where 
\beq
m_K=\frac{4\hbar^2}{J_V\lambda^2}.
\enq
is the effective mass of the kelvons.
Furthermore, from this we see that 
the superfluid density in our case is defined by
\beq
\rho_s(T)=\frac{1}{N_s\lambda}
\sum_{k}\left\{\cos\left(k\lambda/2\right) 
\left[|u_{k}|^2N_{k}+|v_{k}|^2\left(N_{k}+1\right)\right]
-J_V\beta \sin^2\left(k\lambda/2\right)N_{k}\left(N_{k}+1\right)
\right\}.
\enq
Equating the energy cost due to a $4\pi/N_s\lambda$
phase gradient with the thermal energy gives us an estimate for the temperature scale 
$T_\phi=J_V\rho_S(T)\pi^2\lambda/2N_sk_B$
of the phase fluctuations. In order to avoid phase fluctuations, we should be well below 
this temperature scale.

\begin{figure}
\includegraphics[width=0.95\columnwidth]{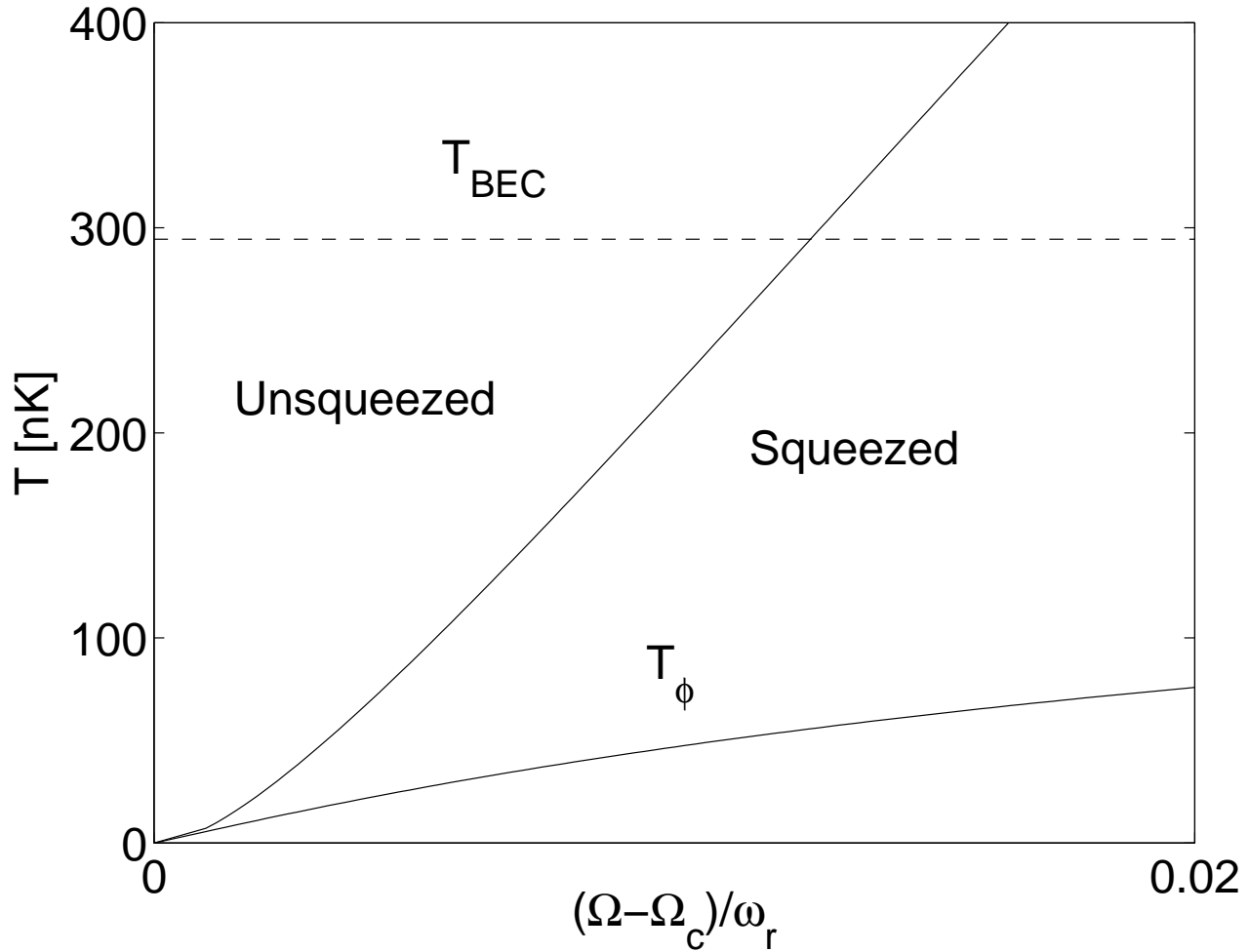}
\vspace{-0.3cm}
\caption[Fig3]{The phase diagram for the vortex squeezing transition. The upper most solid
line is the critical temperature for the squeezing transition. The assumption
of a pure condensate implies temperatures well below the critical temperature of the
Bose-Einstein condensation indicated by the dashed line. Phase fluctuations can be ignored
well below the lowest line.}
\label{fig:phasediag}
\end{figure}

At temperatures higher than $T_\phi$ the vortex can still be squeezed, but the phase fluctuations 
result in different main axis of the vortex position uncertainty ellipse in different 
layers. In Fig.~\ref{fig:phasediag} we plot the critical temperature for Bose-Einstein condensation, 
the critical temperature for the transition into a squeezed vortex, and the temperature 
scale of the phase fluctuations for typical parameter values. This figure indicates 
that the vortex spontaneously squeezes at a temperature that is easily accessible experimentally. 
Below this temperature there is a region in which the vortex is squeezed, but with fluctuating 
main axes. At even lower temperatures the phase fluctuations become negligible 
and long-range order over the size of the system is established. 

\section{Summary and conclusions}
\label{sec:conclusions}
We studied the equilibrium squeezing of the vortex line in an optical lattice
and predicted that strong squeezing is indeed possible in 
the experimentally realistic parameter regime. 
	Although the kelvon interaction induced by 
the vortex displacement is the most important one, other mechanisms for kelvon interactions
also exist. In principle, the kelvons are also 
coupled to the collective modes of the Bose-Einstein condensate, for example, to the 
quadrupole mode~\cite{Martikainen2003b,Martikainen2004a}. 
Such processes typically induce an effective attractive 
kelvon interaction and thus renormalize the value of the interaction 
strength, but do not make it positive. Therefore, we expect that the squeezing 
transition is robust with respect to coupling to the collective modes of the condensate. 

\begin{acknowledgments}
This work is supported by the Stichting voor Fundamenteel Onderzoek der 
Materie (FOM) and by the Nederlandse Organisatie voor 
Wetenschaplijk Onderzoek (NWO).
\end{acknowledgments}

\begin{thebibliography}{32}
\bibitem[{com()}]{comment2004a}
\bibinfo{note}{Alternatively, the incomplete gamma function can be expressed in
  terms of the perhaps more familiar exponential integral as
  $\Gamma[1-p,z]=z^{1-p}E_p(z)$}.
\end{thebibliography}

\end{document}